\documentclass[epj,final]{svjour}
\usepackage{latexsym,amsmath}
\usepackage{graphics}
\usepackage{graphicx}
\usepackage{amssymb}
\usepackage[latin1]{inputenc}
\begin{document}

\title{Surface-Barrier Effects in the Microwave Second-Harmonic
Response of Superconductors in the Mixed State}

\author{A. Agliolo~Gallitto\inst{1} \and G. Giunchi\inst{2} \and M. Li~Vigni\inst{1} \and
G. Vaglica\inst{1}} \institute{INFM and Dipartimento di Scienze
Fisiche e Astronomiche, Universit$\mathrm{\grave{a}}$ di Palermo,
Via Archirafi 36, I-90123 Palermo (Italy)\and EDISON
S.p.A.~-~Divisione Ricerca e Sviluppo, Via U. Bassi 2, I-20159
Milano (Italy)}
%

\abstract{We report on transient effects in the microwave
second-order response of different type of superconductors in the
mixed state. The samples have contemporarily been exposed to a dc
magnetic field, varying with a constant rate of 60 Oe/s, and a
pulsed microwave magnetic field. The time evolution of the signal
radiated at the second-harmonic frequency of the driving field has
been measured for about 500 s from the instant in which the
dc-field sweep has been stopped, with sampling time of $\sim 0.3$
s. We show that the second-harmonic signal exhibits two relaxation
regimes; an initial exponential decay, which endures roughly 10 s,
and a logarithmic decay in the time scale of minutes. Evidence is
given that the decay in the time scale of minutes is ruled by
magnetic relaxation over the surface barrier.
 \PACS{
      {74.25.Ha}{Magnetic properties}\and
      {74.25.Nf}{Response to electromagnetic fields (nuclear magnetic
      resonance, surface impedance, etc.)}\and
      {74.60.Ge}{Flux pinning, flux creep, and flux-line lattice dynamics}
     }
}
\authorrunning{A. Agliolo Gallitto et \textit{al.}}
\titlerunning{Surface-barrier effects in the microwave response}
\maketitle

\section{Introduction}
Measurements of magnetic relaxation in superconductors in the
mixed state allow determining the rate of fluxons to overcome
pinning and surface barriers \cite{Yeshurun}. Indeed, the
interactions of fluxons with both the pinning and the surface
barriers give rise to hysteretic behavior of the magnetization
and, consequently, transient effects. The two barriers affect the
magnetization curve in a different way \cite{Yeshurun}. In
particular, surface-barrier effects manifest themselves in: i)
first-penetration field, $H_p$, higher than the lower critical
field, $H_{c1}$ \cite{Yeshurun,Joseph,deblois}; ii) hysteresis
loop of the magnetization curve asymmetric in the two branches at
increasing and decreasing fields
\cite{Yeshurun,koncikosky,clem,clem2,burla,pissas}; iii) magnetic
relaxation rates different for flux entry and exit
\cite{burla,weir,chikumoto,mishra,rensun}. The effects of the
surface barrier on the magnetization curve is conveniently
investigated at temperatures near $T_c$, where the bulk pinning is
ineffective and, consequently, the asymmetry of the hysteresis
loop due to the surface barrier can be highlighted. On the
contrary, magnetic relaxation over the surface barrier is
conveniently investigated at low temperatures, where the
relaxation over the bulk-pinning potential is expected to occur at
longer times.

In this paper, we investigate the second-order response of
different superconducting samples, exposed to a sweeping dc
magnetic field and a pulsed microwave (mw) magnetic field. It has
been previously shown \cite{claudia,gabriella} that the mw
second-order response of superconductors in the mixed state
exhibits transient effects that, in the time scale of minutes, are
characterized by variation rates different for dc magnetic fields
reached at increasing and decreasing values. It has been
hypothesized that these effects are due to motion of fluxons over
the surface barrier \cite{claudia,gabriella}. The aim of the
present work is to verify the validity of this hypothesis. To this
purpose, we have investigated the time evolution of the mw
response in superconducting samples characterized by the same bulk
properties, but different quality of the surface through which the
magnetic field penetrates. All the measurements have been
performed at the liquid-He temperature, for different values of
the dc field. After the sample has been exposed to a variation of
the dc magnetic field, the signal radiated by the sample at the
second-harmonic (SH) frequency of the driven field exhibits an
initial exponential decay, which lasts about 10 s, and a
logarithmic decay, in the time scale of minutes. The
logarithmic-decay rate depends on the way in which the dc magnetic
field has been reached, i.e. at increasing or decreasing values.
Comparison of the results obtained in samples that differ only for
the quality of the surface through which the magnetic field
penetrates has shown that, in the time scale of minutes, the SH
signal decays slower for smooth surface than for rough surface.
These findings corroborate the hypothesis that, in this time
scale, the time evolution of the SH signal is ruled by magnetic
relaxation over the surface barrier.
\section{Experimental and Samples}
Time evolution of the SH signal has been studied in three samples
of bulk ceramic MgB$_2$, and two samples of Nb polycrystal.

A sample of MgB$_2$ (which we indicate as B$\alpha$) has been
extracted from a pellet sintered from Alfa-Aesar powder at 800°C
in Ar atmosphere, for three hours. It has approximate dimensions
$2\times 1.5 \times 1$ mm$^3$ and $T_c\approx 38$~K. The largest
faces of the sample correspond to the pristine surface of the
pellet from which the sample was extracted; the other faces derive
from the cutting of the pellet with a diamond saw. The faces have
different roughness; those corresponding to the pristine surfaces
of the pellet are smoother than the others.

Other two samples of MgB$_2$ (B1 and B2) have been extracted from
a high-density (2.4~g/cm$^3$) pellet, which has been obtained by
reactive infiltration of liquid Mg on a powdered B preform
\cite{giunchi}. After the reaction in a sealed stainless steel
container, lined with a Nb foil, a thermal treatment has been
performed for two hours in the range of temperatures $850\div
950$~°C. B1 and B2 samples have approximate dimensions $2\times
3\times 0.3$ mm$^3$ and $T_c \approx 39$~K. The largest faces of
the B2 sample have been mechanically polished; they result much
smoother than those of B1 sample.

The two samples of Nb (Nb1 and Nb2) have been extracted from the
same batch, but the largest faces of Nb2 are much smoother than
those of Nb1.

The sample is placed in a bimodal cavity, resonating at the two
angular frequencies $\omega$ and 2$\omega$, with $\omega/2\pi
\approx 3$~GHz, in a region in which the mw magnetic fields
$\textit{\textbf{H}}(\omega)$ and $\textit{\textbf{H}}(2\omega)$
are maximal and parallel to each other. The $\omega$-mode of the
cavity is fed by a pulse oscillator, with pulse width $5~\mu$s and
pulse repetition rate 200 Hz, giving a maximal peak power of
$\approx 50$~W (input peak power of the order of 10 W brings on
microwave magnetic field of the order of 10 Oe in the region of
the cavity in which the sample is located). A low-pass filter at
the input of the cavity cuts any harmonic content of the
oscillator by more than 60 dB. The harmonic signals radiated by
the sample are filtered by a band-pass filter, with more than 60
dB rejection at the fundamental frequency, and are detected by a
superheterodyne receiver. The cavity is placed between the poles
of an electromagnet, which generates dc magnetic fields, $H_0$, up
to $\approx 10$ kOe. All measurements here reported have been
performed at $T = 4.2$~K with $\textit{\textbf{H}}_0\parallel
\textit{\textbf{H}}(\omega)\parallel\textit{\textbf{H}}(2\omega)$.

Before any measurement was performed the sample was zero-field
cooled down to $T = 4.2$~K; $H_0$ was increased up to 10 kOe and
then decreased down to the residual field of the electromagnet.
This preliminary procedure ensures that SH signals arising from
processes occurring in weak links are suppressed by the trapped
flux. The dc field was then swept with a constant rate of $\approx
60$~Oe/s up to fixed values and the evolution of the SH signal was
measured for $\sim 500$~s from the instant in which each $H_0$
value has been reached, with sampling time $\approx 0.3$~s.

Fig.1 shows the time evolution of the SH signal of sample
B$\alpha$, measured from the instant in which $H_0$ has reached
the value of 4 kOe, on increasing (circles) and decreasing
(triangles) the field. Panel (a) shows the results obtained when
both the dc and mw magnetic fields penetrate through the rough
surfaces; panel (b) shows the results obtained when the fields
penetrate through the smooth surfaces. During the first $\sim
10$~s the SH-vs-$t$ curves of panels (a) and (b) show similar
behavior. In this time scale, the SH signal shows an exponential
decay. At longer times, the signal decays following a logarithmic
law; moreover, the decay rate is different for the two
orientations, being smaller when the fields penetrate through the
smooth surfaces. Another peculiarity of the signal decay in the
time scale of minutes concerns the decay rate of the SH signal
after the sweep of the dc field has been stopped: the decay rate
depends on the way the dc field has been reached (at increasing or
decreasing values).
\begin{figure}[t]
\centering
\includegraphics[width=7cm]{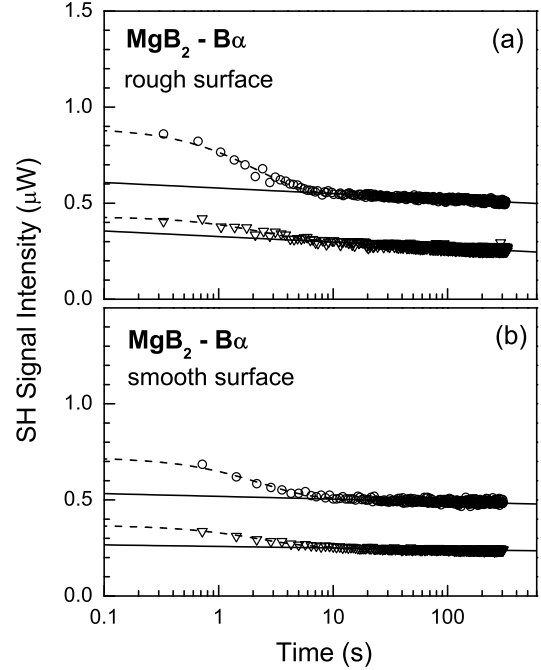}
\caption{Time evolution of the SH signal for the B$\alpha$ sample,
at $H_0$ = 4 kOe, reached on increasing (circles) and decreasing
(triangles) the field. Panels (a) and (b) refer to sample
surfaces, through which the magnetic fields penetrate, of
different roughness. $T = 4.2$~K; input peak power $\approx 2$~W.
Symbols are experimental data; lines are the best-fit curves
obtained as explained in the text.}
\end{figure}

In order to deduce the parameters characteristic of the SH-signal
decay, we have fitted the experimental data by the following
expressions:
\begin{equation}\label{SH(tsec)}
SH = A + B \exp{(-t/\tau)}~~~~0 < t < 10~s
\end{equation}

\begin{equation}\label{SH(tmin)}
SH = C [1 - D \log{(t/t_0)}]~~~~t > 10~s
\end{equation}
\\
with $t_0 = 10$~s and $A = C$ the value that the intensity of the
SH signal takes on at $t=t_0$. \\
The lines in Fig.1 are the best-fit curves.

Measurements performed at different values of the dc magnetic
field allowed us to determine the values of the best-fit
parameters, $\tau$ and $D$, as a function of $H_0$. Fig.2 shows
the field dependence of the characteristic time of the exponential
decay for the B$\alpha$ sample. As one can see, $\tau$ is
independent of the surface roughness and the way the dc magnetic
field is reached; furthermore, it is roughly independent of $H_0$,
within the experimental accuracy.
\begin{figure}[t]
\centering
\includegraphics[width=7cm]{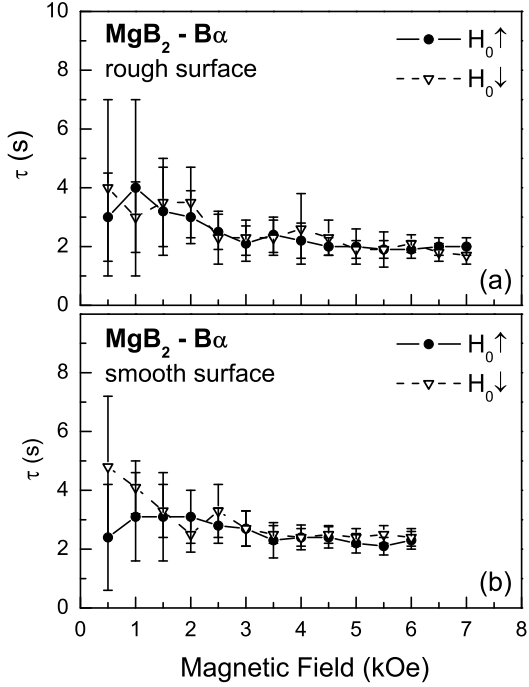}
\caption{Characteristic time of the exponential decay, $\tau$, as
a function of the dc magnetic field, reached at increasing
(circles) and decreasing (triangles) values, for the B$\alpha$
sample. $T$ =~4.2~K; input peak power $\approx 2$~W. Lines are
leads for eyes.}
\end{figure}

Fig.3 shows the value of the best-fit parameter $D$ for the
B$\alpha$ sample as a function of $H_0$. The rate of the
logarithmic decay depends on $H_0$ and its sweep direction; it is
larger for magnetic fields reached at decreasing values than for
fields reached at increasing values. On increasing $H_0$, the
logarithmic-decay rates for negative and positive field variations
approach each other. Furthermore, comparison between the results
of panels (a) and (b) shows that the logarithmic-decay rate of the
SH signal is about two times smaller when the magnetic fields
penetrate through the smoother surface, in all the range of fields
investigated.
\begin{figure}[tb]
\centering
\includegraphics[width=7cm]{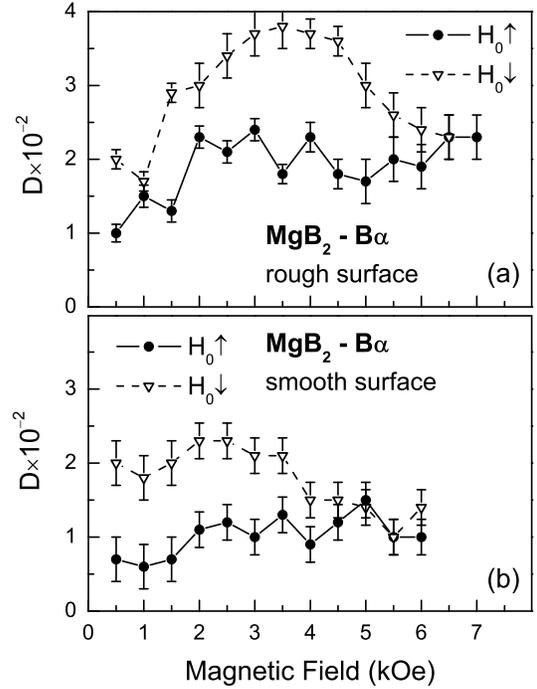}
\caption{Field dependence of the logarithmic-decay rate, $D$, for
the B$\alpha$ sample. Circles and triangles refer to measurements
performed at $H_0$ values reached at increasing and decreasing
fields, respectively. Lines are leads for eyes}
\end{figure}

Fig.4 shows the time evolution of the SH signal for B1 (a) and B2
(b) samples, at $H_0 = 5$~kOe reached on increasing (circles) and
decreasing (triangles) the field. It is worth to remember that B1
and B2 samples have the same bulk properties, but the faces of B2
sample have been polished and are much smoother than those of B1
sample. Comparison between the results of panels (a) and (b) shows
that in the sample with polished faces the SH signal is roughly
steady, in the time scale investigated. In the inset of panel (a)
we report the field dependence of the logarithmic-decay rate of
the SH signal, obtained by fitting the SH-vs-$t$ curves of B1
sample by Eq.(2). By fitting the initial decay of the SH signal of
B1 sample by Eq.(1) we have obtained values of the best-fit
parameter $\tau$ close to those obtained in the B$\alpha$ sample
(see Fig.2).
\begin{figure}[t]
\centering
\includegraphics[width=7cm]{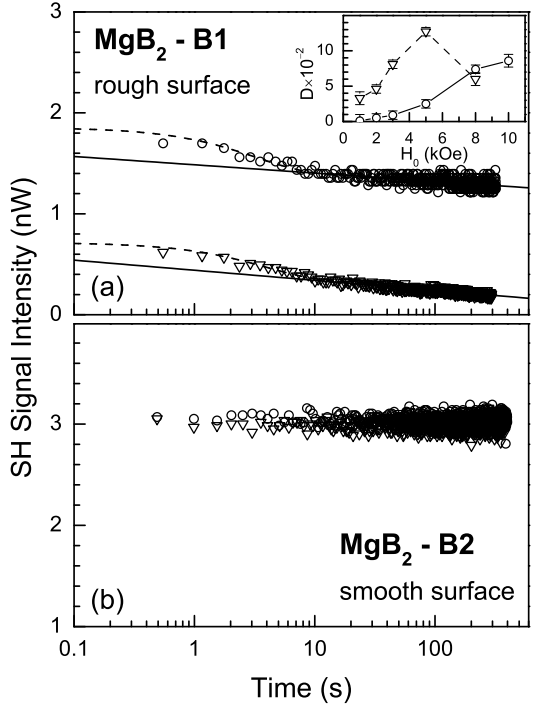}
\caption{Time evolution of the SH signal for B1 (a) and B2 (b)
samples, at $H_0$ = 5 kOe, reached on increasing (circles) and
decreasing (triangles) the field. $T$ = 4.2 K; input peak power
$\approx 4$~W. Symbols are experimental data; lines are the
best-fit curves obtained as explained in the text. Inset of panel
(a) shows the field dependence of the best-fit parameter $D$
(lines are leads for eyes).}
\end{figure}

In Fig.5 we show the time evolution of the SH signal for Nb1 (a)
and Nb2 (b) samples, at $H_0 = 4$~kOe, reached on increasing
(circles) and decreasing (triangles) the field. It is worth to
remember that the two samples differ for the quality of the
surfaces, as indicated in the figure. Again, the SH signal of the
sample with smooth surface decays slower than that of the sample
with rough surface.
\begin{figure}[h]
\centering
\includegraphics[width=7cm]{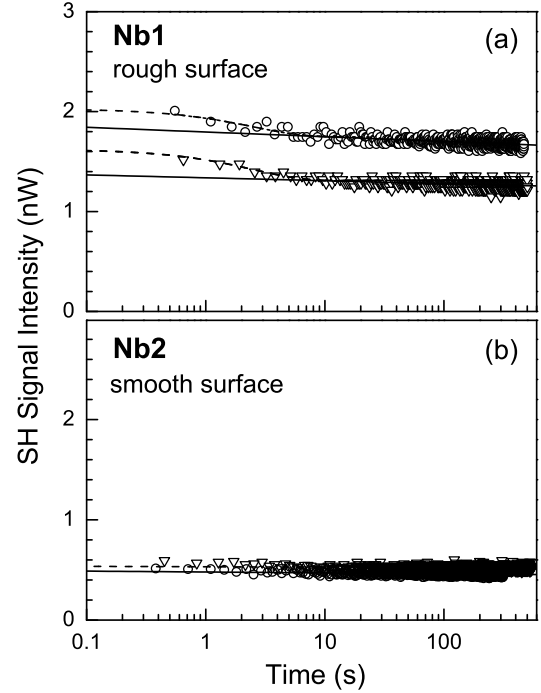}
\caption{Time evolution of the SH signal for Nb1 (a) and Nb2 (b)
samples, at $H_0$ = 4 kOe, reached on increasing (circles) and
decreasing (triangles) the field. $T $= 4.2 K; input peak power
$\approx 4$~W. Symbols are experimental data; lines are the
best-fit curves obtained as explained in the text.}
\end{figure}

By fitting the SH-vs-$t$ curves of Nb samples, we have obtained
values of the best-fit parameter $D$ one order of magnitude
smaller than those obtained in the MgB$_2$ samples. Due to the
slow variation, the logarithmic-decay rate of the SH signal in the
Nb samples can be determined only with a large uncertainty. This
finding does not allow comparing quantitatively the results
obtained for $H_0$ reached at increasing and decreasing fields.
Fig.6 shows a comparison between the $D$ values obtained in the
two Nb samples, at increasing (a) and decreasing (b) fields.

A further peculiarity of the SH response, which can be seen in the
figures, is its hysteretic behavior. As one can see, for the
samples in which the SH signal shows a noticeable time evolution,
the intensity of the signal is different for $H_0$ reached at
increasing and decreasing fields. We infer that the hysteresis is
strictly related to the transient effects.

Measurements performed in the MgB$_2$ samples at different
temperatures have shown that, up to few K below $T_c$, the
peculiarities of the SH signal are not significantly affected by
the temperature; in particular, $\tau$ and $D$ take on values of
the same order of magnitude as those obtained at $T$ = 4.2 K. Only
at temperatures very close to $T_c$ the SH signal is stationary
and no hysteresis is observed.
\begin{figure}[lb]
\centering
\includegraphics[width=5.5cm]{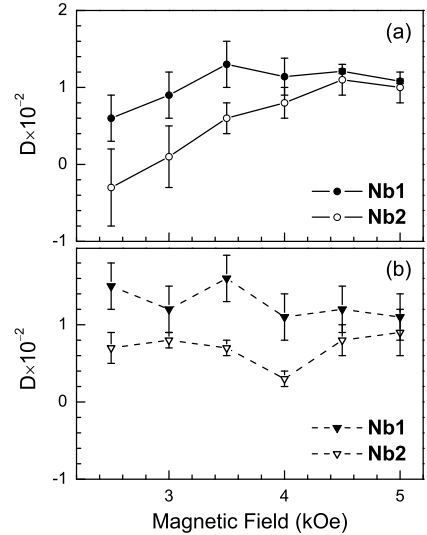}
\caption{Logarithmic decay rate, $D$, of the SH signal of the two
Nb samples, as a function of $H_0$ reached at increasing (a) and
decreasing (b) fields. Lines are leads for eyes.}
\end{figure}
\section{Discussion}
Nonlinear electromagnetic response of both conventional and
high-$T_c$ superconductors has been discussed by different authors
\cite{samoilova,jeffries,tinkham,muller,noi161,bean,noi159,noiPRB,agliolo,misha}.
Several mechanisms give rise to emission of signals oscillating at
harmonic frequencies of the driving field. At low magnetic fields
and low temperatures, harmonic generation has been ascribed to
nonlinear processes in weak links, impurities and
intergrain-fluxon dynamics \cite{jeffries,tinkham,muller,noi161}.
At magnetic fields higher than the lower critical field, when the
weak links are decoupled, harmonic emission has been ascribed to
intragrain-fluxon dynamics \cite{bean,noi159}. At temperatures
near $T_c$, modulation of the order parameter by the em field is
the main source of nonlinearity \cite{noiPRB,agliolo,misha}. To
our knowledge, none of the models reported up to now in the
literature discuss relaxation phenomena in the em response.

At low temperatures, after the samples have been exposed to
magnetic fields larger than the first-penetration field, it is
reasonable to hypothesize that a critical state develops.
Nonlinear magnetization of superconductors in the critical state
has been for the first time studied by Bean \cite{bean}. The Bean
model accounts quite well for the nonlinear response of
conventional superconductors to low-frequency em fields. It is
based on the hypothesis that the critical current does not depend
on the magnetic field; furthermore, it tacitly assumes that the
fluxon lattice follows adiabatically the em field variations. On
these hypotheses, the response of the sample is even during the
period of the em field; consequently, only odd-harmonic emission
is expected. Even harmonics can be expected by taking into account
the field dependence of the critical current
\cite{tinkham,muller}, according to the Anderson and Kim critical
state model \cite{kim}. However, for $H_{dc} \gg H_{ac}$ the
results obtained by using the Anderson-Kim model converge to those
of Bean and only odd harmonics are expected \cite{tinkham}.
Nevertheless, it has been reported that superconductors in the
critical state exposed to pulsed mw fields exhibit odd as well as
even-harmonic emission \cite{noi159}, even when $H_{dc} \gg
H_{ac}$. The mw SH emission by superconductors in the critical
state has been discussed by Ciccarello et $al$. \cite{noi159}, who
have elaborated a phenomenological model, based on the Bean model,
in which the additional hypothesis is put forward that
superconductors in the critical state operate a rectification
process of the mw input field. The process arises because of the
inertia of the fluxon lattice to follow high-frequency fields. It
has been supposed that, due to the rigidity of the fluxon lattice,
the induction field inside the sample does not follow
adiabatically the mw field variations, except when the variations
bring about motion of fluxons in the surface layers of the sample.
From this model, it is expected that superconductors in the
critical state radiate stationary SH signals, whose intensity is
independent of the magnetic-field-sweep direction.

In the framework of the above-mentioned models, time evolution of
harmonic signals could arise if the hypothesis is done that the
critical state evolves toward a thermal-equilibrium state, by flux
creep. However, several experimental evidences disagree with this
hypothesis. Our results show that the time evolution of the SH
signal exhibits two distinct regimes of relaxation: i) an
exponential decay, with characteristic time independent of the
surface roughness and the magnetic-field-sweep direction; ii) a
logarithmic decay, with characteristic rate that strongly depends
on the surface roughness and the field-sweep direction. The
exponential decay is characterized by times of the order of
seconds; so, it cannot be ascribed to magnetic relaxation by creep
processes, especially so at low temperatures. On the other hand,
relaxation through the pinning barrier does not justify the
different decay rates observed after increasing and decreasing
fields for times of the order of minutes. The dependence of the
decay rate on the roughness of the sample surface, through which
the magnetic field penetrates, strongly supports the idea that the
decay of the SH signal in this time scale is due to
surface-barrier effects.

In spite of the clear experimental evidence that the logarithmic
decay of the SH signal is ruled by magnetic relaxation over the
surface barrier, up to now there is not any model to
quantitatively justify the experimental results. Indeed, it is
worth noting that neither the Bean model nor the Ciccarello model
take into account the presence of the surface barrier. Although in
the literature are reported many studies on the surface barrier,
none of them concern the effects of the barrier on the nonlinear
em response. It would be of great importance to understand if
surface-barrier effects induce relaxation of the signal or they
could be themselves source of second-order response. As shown by
Clem \cite{clem,clem2}, the main effect of the surface barrier is
the asymmetric response of fluxons for increasing and decreasing
fields. We suggest that, due to this asymmetry, the presence of
the surface barrier may be source of SH emission. In order to have
this effect the amplitude of the oscillating field should be high
enough to nucleate (or annihilate) fluxons near the sample surface
during the em-field period. In this case, time evolution of the SH
signal can be expected because after the magnetic-field sweep is
stopped the amplitude of the surface barrier increases. The
variation rate of the SH signal should be related to the rate of
variation of the surface barrier that, as suggested by Burlachkov
\cite{burla}, is expected different for flux entry and exit.

Our results rule out the possibility that the initial exponential
decay of the SH signal is ascribable to surface-barrier effects.
Magnetic relaxation having characteristic times of the order of
seconds has been reported by different authors
\cite{zola,kalinsky}. The authors of Ref.\cite{zola} have
discussed the exponential decay of the dc magnetization, detected
during the first $\sim10$ s of the relaxation process in
BSCCO(2223)/Ag tapes, considering diffusive motion of fluxons
induced by the variation of the dc magnetic field. Kalisky et
$al$. \cite{kalinsky} have reported several studies on the time
evolution of the magnetic-induction field in BSCCO crystals. They
show that, during the magnetic-field sweep, two distinct vortex
states coexist, characterized by different values of the
persistent current density ("high" and "low"). The
high-persistent-current state has been defined as a transient
disordered vortex state (TDVS), because it decays with time when
the external field is kept constant. For increasing fields, the
TDVS state is located near the sample edges; for decreasing
fields, it is located in the interior of the sample. When the
magnetic field is kept constant, the "break" between the two
states moves with time toward the sample edges or center,
dependently on the way the magnetic field has been reached (at
increasing or decreasing values). Though the flux configuration is
different at increasing and decreasing fields, the characteristic
time with which the TDVS evolves toward an ordered state does not
depend on the magnetic-field-sweep direction; furthermore, it
results of the order of seconds.

We think that the exponential decay of the SH signal, revealed in
the time scale of seconds, is ascribable to processes similar to
those discussed in Refs.\cite{zola,kalinsky}. In particular, we
suggest that it is related to the following process. During the
field sweep, the fluxons arrange themselves in a configuration
incompatible with the critical state; as soon as the field sweep
is stopped, a diffusive motion of fluxons sets in; the process
ends when the flux density in the bulk reaches the appropriate
value for the critical state.
\section{Conclusions}
We have reported on transient effects in the mw second-order
response of different types of superconductors in the mixed state.
The measurements have been performed in a time window of $\sim
500$~s, after exposing the sample to a sweeping dc magnetic field
ranging from 0 to 10 kOe. We have shown that during the first
seconds the SH signal decays exponentially, while in the time
scale of minutes it shows a logarithmic decay. The characteristic
time of the initial decay does not depend on the field-sweep
direction and the roughness of the sample surface through which
the magnetic field penetrates. On the contrary, the variation rate
of the logarithmic decay strongly depends on the surface roughness
and the way in which the field is reached. These findings provide
evidence that the two regimes of decay arise from different
processes. In particular, we have suggested that the initial decay
is related to diffusive motion of fluxon, which occurs during the
time in which the critical state develops, while the logarithmic
decay arises from magnetic relaxation over the surface barrier.
Further investigation is necessary to understand whether
surface-barrier effects are source of the second-order response,
or fluxon motion over the surface barrier induces relaxation of SH
signals arising from known nonlinear processes.
\section{Acknowledgements} The authors are very glad to thank I. Ciccarello
for critical reeding of the manuscript and helpful suggestions; G.
Lapis and G. Napoli for technical assistance.

\begin{flushright}\today \end{flushright}


\begin{thebibliography}{99}

\bibitem{Yeshurun}Y.~Yeshurun, A.~P.~Malozemoff, A.~Shaulov, Rev. Mod. Phys. {\bf 68}, (1996)
 911 and references therein.

\bibitem{Joseph}A.~S.~Joseph, W.~J.~Tomasch, Phys. Rev. Lett. {\bf
12}, (1964) 219.

\bibitem{deblois}R.~W.~De Blois, W.~De~Sorbo, Phys. Rev. Lett. {\bf 12},
(1964) 499.

\bibitem{koncikosky}M.~Konczykowski, L.~Burlachkov, Y.~Yeshurun,
F.~Holtzberg, Phys. Rev. {\bf B 43}, (1991) 13707.

\bibitem{clem}J.~R.~Clem, in {\sl Proceedings of the LTP$13^{th}$ Conference},
edited by K.~D.~Timmerhaus, W.~J.~O'Sullivan, and E.~F.~Hammel
(Plenum, New York 1974) vol.{\bf 3}, 102.

\bibitem{clem2}J.~R.~Clem, J. Appl. Phys. {\bf 50}, (1979) 3518.

\bibitem{burla}L. Burlachkov, Phys. Rev. {\bf B 47}, (1993) 8056.

\bibitem{pissas}M. Pissas, E. Moraitakis, D.~Stamopoulos, G.~Papavassiliou,
V.~Psycharis, S.~Koutandos, J. Supercond. {\bf 14}, (2001) 615.

\bibitem{weir}S. T. Weir, W. J. Nellis, Y.~Dalichauch, B.~Lee,
M.~B.~Maple, J.~Z.~Liu, R.~N.~Shelton, Phys. Rev. {\bf B 43},
(1991) 3034.

\bibitem{chikumoto}N. Chikumoto, M. Konczykowski, N.~Motohira,
A.~P.~Malozemoff, Phys. Rev. Lett. {\bf 69}, (1992) 1260.

\bibitem{mishra}P. K. Mishra, G. Ravikumar, T.~V.~Chandrasekhar
Rao, V.~C.~Sahni, S.~S.~Banerjee, S.~Ramakhrishnan, A.~K.~Grover,
M.~J.~Higgins, Physica {\bf C 340}, (2000) 65.

\bibitem{rensun}Yang Ren Sun et al., Phys. Rev. {\bf B 50}, (1994)
3330.

\bibitem{claudia}A. Agliolo Gallitto, I.~Ciccarello, C.~Coronnello,
M.~Li~Vigni, Physica {\bf C 402}, (2004) 309.

\bibitem{gabriella}A. Agliolo Gallitto, M. Li Vigni, G.~Vaglica,
Physica {\bf C 404}, (2004) 6.

\bibitem{giunchi}G. Giunchi, Int. J. Mod. Phys. {\bf B 17}, (2003)
453.

\bibitem{samoilova}T. B. Samoilova, Supercond. Sci. Technol. {\bf
8}, (1995) 259 and references therein.

\bibitem{jeffries}Q. H. Lam, C. D. Jeffries, Phys. Rev. {\bf B 39}, (1989) 4772.

\bibitem{tinkham}L. Ji, R.~H.~Sohn, G.~C.~Spalding, C.~J.~Lobb, M.~Tinkham, Phys. Rev.
{\bf B 40}, (1989) 10936.

\bibitem{muller}K. H. Muller, J. C. MacFarlane, R.~Driver, Physica
{\bf C 158}, (1989) 69.

\bibitem{noi161} I. Ciccarello, M. Guccione, M.~Li~Vigni, Physica
{\bf C 161}, (1989) 39.

\bibitem{bean}C. P. Bean, Rev. Mod. Phys. {\bf 36}, (1964) 31.

\bibitem{noi159}I. Ciccarello, C. Fazio, M.~Guccione, M.~Li~Vigni, Physica
{\bf C 159}, (1989) 769.

\bibitem {noiPRB}I. Ciccarello, C. Fazio, M.~Guccione, M.~Li Vigni, M.~R.~Trunin, Phys. Rev. {\bf
B 49}, (1994) 6280.

\bibitem{agliolo}A. Agliolo Gallitto, M. Li Vigni, Physica {\bf C
305}, (1998) 75.

\bibitem{misha}M. R. Trunin, G. I. Leviev, J. Phys. III (France) {\bf
2}, (1992) 355.

\bibitem{kim}Y. B. Kim, C. F. Hempstead, A. R. Strnad, Phys. Rev. Lett.
{\bf 9} (1962) 306.

\bibitem{zola}D. Zola, M. Polichetti, S.~Pace, Int. J. Mod. Phys.
{\bf B 14}, (2000) 2890.

\bibitem{kalinsky}B. Kalisky, Y. Bruckental, A.~Shaulov, Y.~Yeshurun, Phys. Rev.
{\bf B 68}, (2003) 224515.
\end{thebibliography}
\end{document}